\documentclass[10pt,a4paper]{article}
\usepackage[a4paper,left=2.5cm,right=2.5cm,top=2.5cm,bottom=2.5cm]{geometry}
\usepackage[super,compress]{cite}
\usepackage{graphicx}
\usepackage{authblk}
\usepackage{xcolor}
\usepackage{subcaption}

\usepackage{amsmath}

\DeclareMathOperator{\sech}{sech}

\begin{document}

\title{Analytical calculation of the observational parameters  for tachyon inflation}


\author[1]{Marko Stojanovic\thanks{marko.stojanovic@pmf.edu.rs}}
\author[2]{Neven Bili\'c\thanks{bilic@irb.hr}}
\author[3]{Goran S. Djordjevic\thanks{gorandj@junis.ni.ac.rs}}
\author[3]{Dragoljub D. Dimitrijevic\thanks{ddrag@pmf.ni.ac.rs}}
\author[3]{Milan Milosevic\thanks{milan.milosevic@pmf.edu.rs}}

\affil[1]{Faculty of Medicine, University of Ni\v s, Ni\v s, Serbia}

\affil[2]{Division of Theoretical Physics, Rudjer Bo\v{s}kovi\'{c} Institute, Zagreb, Croatia}
\affil[3]{Department of Physics, University of Ni\v s, Ni\v s, Serbia}

\maketitle

\begin{abstract}
We investigate the possibility of analytically calculating observational parameters in tachyon inflation cosmology, using the Hubble expansion rate as a function of the tachyon field. First, in light of the newer Planck results, we analyze previous investigations in which the test Hubble rate functions were confronted with Planck 2013 data. We propose and analyze a number of new test Hubble rate functions, finding  considerable improvement and approaching reasonable agreement with recent observational data. For a class of these functions, we were able to carry out analytical calculations. As a novelty, we introduce a functional dependence of the slow-roll Hubble flow parameters as an input, which facilitates an analytical study of inflation. In this way, we obtain explicit expressions for the scalar spectral index ($n_{\rm s}$) and the tensor-to-scalar ratio ($r$). We confront our analytical and numerical predictions with Planck observational constraints. Finally, we discuss the proposed linear dependence between the first two slow-roll parameters in light of recent data from ACT DR6, DESI, and others, as well as the potential to generalize this dependence to other inflation models.
\end{abstract}


\section{Introduction}

The idea of cosmological inflation, supported by observational data, is the best candidate to overcome some problems in the early universe and explain the origins of large-scale structures [\citen{Guth:1980zm,Odintsov:2023weg}]. The most common way to realize the inflationary epoch is to introduce a single (scalar) field dubbed the inflaton. One relevant possibility is inflation driven by the tachyon scalar field, motivated by the string or M-theory [\citen{Sen:2002in,Sen:2002nu,sen3}]. It was shown that the process of tachyon condensation might provide accelerated expansion of the universe required for inflation. Effective field theory can describe that process using a nonstandard Dirac–Born–Infeld (DBI) form of the Lagrangian. Many papers related to tachyon inflation have appeared [\citen{Gibbons:2002md,Feinstein:2002aj,Fairbairn:2002yp,Sami:2002fs,Kofman:2002rh,Frolov:2002rr,Steer:2003yu,Gibbons:2003gb}]. It is worth noting that tachyon inflation suffers from the reheating problem since the conventional reheating mechanism in the tachyon case does not work. The interesting scenario to overcome this issue was proposed in Ref. [\citen{Bilic:2017yll}].
 
The common feature of inflationary models is the slow-roll approximation [\citen{Liddle:1994dx}]. In that scenario, inflation is governed by the slow evolution of the inflaton field potential, which is specified phenomenologically. Many inflationary models have been proposed in the literature, using an alternative approach based on the Hubble expansion rate as a function of the inflaton field $\phi$. This approach is often referred to as the ``Hamilton-Jacobi formalism" [\citen{liddle,Steer:2003yu,lyth}]. It allows one to consider the Hubble rate $H(\phi)$, rather than the potential $V(\phi)$, as the fundamental quantity to be specified. Because $H$, unlike $V$, is a geometric quantity, inflation is described more naturally in that language [\citen{lyth,Videla:2016ypa}]. In Ref. [\citen{Aghamohammadi:2014aca}], the inflation model with a tachyon field minimally coupled to gravity has been investigated. The Hamilton-Jacobi formalism applied to a warm inflationary scenario is studied in Refs. [\citen{Akhtari:2017mxc, Sayar:2017pam}] and formulated for inflation with non-minimal derivative coupling in Ref. [\citen{Sheikhahmadi:2016wyz}]. This paper also compares the non-minimal derivative coupling with the minimal coupling case. The Hamilton-Jacobi formalism has recently been used in a model of $k$-inflation [\citen{Yang:2023rjt}], in the generalized Chaplygin gas model [\citen{Ignatov:2021mpo}], and in a model of anisotropic inflation [\citen{Cicciarella:2019ihh}]. The tachyon field as a model for inflation has been extensively studied in the literature using the slow-roll approximation [\citen{Dimitrijevic:2018nlc,Bilic:2017orf,Bilic:2016fgp}]. A detailed study of inflation with tachyon field minimally coupled to gravity has been carried out in [\citen{Steer:2003yu}]. We shall refer to this type of models as the ``standard tachyon inflation".
In our previous studies, we have investigated various aspects and the phenomenology of tachyon inflation [\citen{Bilic:2025rjg,Stojanovic:2023dni,Stojanovic:2023qgm,Bilic:2018uqx}]. In the current work, we first revisit standard tachyon inflation via the Hamilton-Jacobi approach proposed in Ref. [\citen{Aghamohammadi:2014aca}], incorporating a novel dependence of the Hubble expansion rate on the scalar tachyon field. We obtain their results for the scalar spectral index ($n_{\rm s}$) and the tensor-to-scalar ratio ($r$), which were in reasonable agreement with the Planck 2013 data  but not quite in agreement with the latest Planck data. In the next step, we consider various functional forms of the Hubble rate, find the theoretical and numerical values of the parameters $n_{\rm s}$ and $r$, and propose a generalization. Finally, we confront our results with observations of Planck 2018 [\citen{Planck:2018jri}], ACT DR6 [\citen{act}], and DESI [\citen{desi}].

The remainder of the paper is organized as follows. In section \ref{dynamics}, we provide a brief recapitulation of the tachyon dynamics in the standard cosmology. In the following section, section \ref{hubble}, we review the Hamilton-Jacobi formalism for inflation. In section \ref{various}, we calculate the slow roll and observational parameters for the assumed Hubble rate functions and confront the results with observational data. In section \ref{method}, we propose a new method for generating Hubble rate functions that allows analytical calculation of the observational parameters. In the concluding section, section \ref{conclude}, we summarize our results and propose future research.
\section{Dynamics of the tachyon inflation}
\label{dynamics}

Consider the action for a tachyon field $\theta$ minimally coupled to gravity is (see, e.g., [\citen{sen3}])
\begin{equation}
S=\int d^4 x\sqrt{-g}\left(\frac{M_{\rm Pl}^2}{2}R-V(\theta)\sqrt{1-g^{\mu\nu}\theta_{,\mu}\theta_{,\nu}}\right),
\end{equation}
where $M_{\rm Pl}=(8\pi G_{\rm N})^{-1/2}$ is the reduced Planck mass, $R$ is the Ricci scalar and $V(\theta)$ is a potential. In the flat, homogeneous, and isotropic universe, described by the FLRW metric
\begin{equation}
ds^2=dt^2-a^2(t)(dx^2+dy^2+dz^2),
\end{equation}
the Friedmann equations are of the form
\begin{equation}
H^2=\frac{1}{3M_{\rm Pl}^{2}}\rho,
\label{F1}
\end{equation}
\begin{equation}
\dot{H}=-\frac{1}{2M_{\rm Pl}^{2}}(\rho+p),
\label{F2}
\end{equation}
where the overdot indicates a derivative with respect to the cosmic time $t$. The Hubble expansion rate is defined as $H(t)\equiv\dot{a}/a$. The energy conservation yields
\begin{equation}
\dot{\rho}=-3H(p+\rho).
\label{dotrho}
\end{equation}
The spatially homogenous tachyon field $\theta(t)$ describes an ideal cosmological fluid with pressure $p$ and energy density $\rho$ defined as
\begin{equation}
p=-V\sqrt{1-\dot{\theta}^2},
\label{eqp}
\end{equation}
\begin{equation}
\rho=\frac{V}{\sqrt{1-\dot{\theta}^2}}.
\label{eqrho}
\end{equation}
The model is fixed by specifying the tachyon potential. A wide range of potentials motivated by string theory can be used. From (\ref{F1}), (\ref{F2}), (\ref{eqp}), and (\ref{eqrho}), one obtains the time dependence of the tachyon field in terms of the Hubble rate and its derivative
\begin{equation}
\dot{\theta}^2=-\frac{2}{3}\frac{\dot{H}}{H^2}.
\label{dottheta}
\end{equation}
For comparison, in the canonical inflation with a single scalar field $\phi$, one finds the expression [\citen{Liddle:1994dx}]
\begin{equation}
\dot{\phi}^2=-2M_{\rm Pl}^2\dot{H}.
\end{equation}
In a model for inflation with non-minimal derivative coupling, as formulated in Ref. [\citen{Sheikhahmadi:2016wyz}], equation (\ref{dottheta}) takes the same form.

Equation (\ref{dotrho}), with the help of (\ref{eqp}) and (\ref{eqrho}), gives a second order equation 
\begin{equation}
\frac{\ddot{\theta}}{1-\dot{\theta}^2}+3H\dot{\theta}+\frac{V_{,\theta}}{V}=0,
\label{eqddottheta}
\end{equation}
where the subscript $,\theta$ denotes a derivative with respect to $\theta$.

It is convenient to introduce and exploit the hierarchical slow-roll parameters defined as [\citen{Steer:2003yu}]
\begin{equation}
\varepsilon_{0}\equiv \frac{H_{*}}{H},
\end{equation}
\begin{equation}
\varepsilon_{i+1}\equiv \frac{d\ln|\varepsilon_{i}|}{dN},\quad i\geq 0,
\label{slparametersdef}
\end{equation}
where $H_{*}$ is the Hubble rate at some chosen time.
The first three parameters are
\begin{equation}
\varepsilon_{1}=-\frac{\dot{H}}{H^2},
\label{defe1}
\end{equation}
\begin{equation}
\varepsilon_{2}=\frac{\dot{\varepsilon}_{1}}{\varepsilon_{1} H},
\label{defe2}
\end{equation}
\begin{equation}
\varepsilon_{3}=\frac{\dot{\varepsilon}_{2}}{\varepsilon_{2} H}.
\label{defe3}
\end{equation}
During the slow-roll regime, the slow-roll parameters are much less than one ($\varepsilon_{i}\ll 1$).

Using the expansion rate $H$, one defines the so-called e-fold number $N$ via
\begin{equation}
H=\frac{dN}{dt}
\label{efold0}
\end{equation}
and the number of efolds at the end of inflation 
\begin{equation}
N_*\equiv\int_{t_{\rm i}}^{t_{\rm f}}H dt,
\label{efold}
\end{equation} 
where the subscripts $\rm i$ and $\rm f$ denote the beginning and the end of inflation, respectively. We will adopt the widely accepted assumption that the number of e-folds at the end of inflation is approximately $N_*\simeq 60$ to resolve the notorious cosmological problems, such as the horizon and flatness problems.

In the slow-roll regime, we have
\begin{equation}
\ddot{\theta}\ll 3H\dot{\theta},\quad \dot{\theta}^2\ll 1.
\label{eq0002}
\end{equation}
Inflation ends when the first slow-roll parameter exceeds unity, i.e., at the time when $\varepsilon_{1}(t_{\rm f})=1$. By imposing the slow-roll conditions (\ref{eq0002}), we may neglect $\dot{\theta}^2$ in equation (\ref{eqrho}) and the term proportional to $\ddot{\theta}$ in equation (\ref{eqddottheta}). Within this approximation, one finds [\citen{Fairbairn:2002yp}]
\begin{equation}
H^2\simeq \frac{V}{3M_{\rm Pl}^2},
\label{sreqs}
\end{equation}
\begin{equation}
\dot{\theta}\simeq -\frac{V_{,\theta}}{3HV}.
\end{equation}

\section{Hubble rate as a function of the tachyon field}
\label{hubble}

In this section, we briefly describe a formalism based on the idea that the tachyon field can serve as a substitute for the time variable when the scalar field varies monotonically. Using
\begin{equation}
\dot{H}=H_{,\theta}\dot{\theta},
\label{dothdottheta}
\end{equation}
the Friedmann equations (\ref{F1}) and (\ref{F2}) can be rewritten as
\begin{equation}
H_{,\theta}^2-\frac{9}{4}H^4+\frac{1}{4M_{\rm Pl}^2}V^2=0,
\label{HJ1}
\end{equation}
\begin{equation}
H_{,\theta}=-\frac{3}{2}H^2\dot{\theta}.
\label{HJ2}
\end{equation}
It was noted [\citen{Steer:2003yu,lyth}] that equation (\ref{HJ1}) resembles the Hamilton-Jacobi equation.
So, in the following, we shall refer to equations (\ref{HJ1}) and (\ref{HJ1}) as the Hamilton-Jacobi system.
As a consequence of (\ref{dottheta}) and (\ref{HJ2}) the field inceases (decreases) over time if $H_{,\theta}<0$ ($H_{,\theta}>0$).
The advantage of the Hamilton-Jacobi approach is that the potential can be expressed in terms of the Hubble rate.
From equations (\ref{F1}), (\ref{eq0002}) and (\ref{HJ2}) it follows
\begin{equation}
V(\theta)=3M_{\rm Pl}^{2}H^2\sqrt{1-\frac{4}{9}\frac{H_{,\theta}^2}{H^4}}.
\label{potential}
\end{equation}
However, as pointed out in [\citen{Feinstein:2002aj}], the main drawback of this procedure is that it may often lead to an unphysical potential.

By combining (\ref{HJ2}) with (\ref{defe1}) and (\ref{defe2}), we obtain the Hubble flow parameters in terms of $H$ and its derivatives with respect to the field
\begin{equation}
\varepsilon_{1}=\frac{2}{3}\frac{H_{,\theta}^2}{H^4},
\label{eqHJe1}
\end{equation}
\begin{equation}
\varepsilon_{2}=\frac{4}{3}\frac{1}{H^4}\left(2H_{,\theta}^2-H_{,\theta\theta}H\right).
\label{eqHJe2}
\end{equation}
Another important expression derived from (\ref{efold}) and (\ref{HJ2}) is
\begin{equation}
N_*=-\frac{3}{2}\int_{\theta_{\rm i}}^{\theta_{\rm f}}\frac{H^3}{H_{,\theta}}d\theta.
\label{NHJ}
\end{equation}
Given $H(\theta)$, integrating Eq. (\ref{HJ2}) yields $\theta(t)$. Then, the expansion scale can be calculated using
\begin{equation}
a(t)=a_{0} \exp \int_{0}^{t}H(\theta(t^\prime))dt^\prime.
\label{sf}
\end{equation}
As already pointed out in [\citen{Djordjevic:2024fxy}], the Hamilton–Jacobi approach is also suitable for studying the attractor behavior of the inflationary models.

\subsection{The scalar spectral index and the tensor-to-scalar ratio}\label{sec4}
It is well known that in the standard inflation, including tachyon inflation, the scalar spectral index $n_{\rm s}$ and the tensor-to-scalar ratio $r$ in the lowest order in the slow-roll parameters are given by [\citen{Steer:2003yu}]
\begin{equation}
n_{\rm s}=1-2\varepsilon_{1}-\varepsilon_{2},
\label{defns}
\end{equation}
\begin{equation}
r=16\varepsilon_{1},
\label{defr}
\end{equation}
where the values of $\varepsilon_{1}$ and $\varepsilon_{2}$ are taken at the beginning of inflation when $\theta_{\rm i}\equiv \theta(t_{\rm i})$, i.e. $\varepsilon_{1}\equiv \varepsilon_{1}(\theta_{\rm i})$ and $\varepsilon_{2}\equiv \varepsilon_{2}(\theta_{\rm i})$. A model of inflation may be considered viable if the values of $n_{\rm s}$ and $r$ are consistent with observational data.

Using the standard approach, the parameters $n_{\rm s}$ and $r$ can be calculated analytically as follows:
\begin{itemize}
\item Given $H(\theta)$, compute the slow-roll parameters $\varepsilon_{1}$ and $\varepsilon_{2}$ as functions of the field $\theta$.
\item Compute the value of the field at the end of inflation $\theta_{\rm f}$ using the condition $\varepsilon_{1}(\theta_{\rm f})=1$.
\item Determine the value of the field at the beginning of inflation $\theta_{\rm i}$ from the differential equation for the e-fold number of $N$ (derived from (\ref{NHJ}))
\begin{equation}
2H_{,\theta}dN=-3H^3d\theta,
\end{equation}
for a given value $N_{*}\equiv N_{\rm f}-N_{\rm i}$. In our calculations we take $N_{*}$ in the interval $55\leq N_{*}\leq 65$.
\item Use $\theta_{\rm i}$ to find $\varepsilon_{1}(\theta_{\rm i})$ and $\varepsilon_{2}(\theta_{\rm i})$, and calculate $n_{\rm s}$ and $r$ using Eqs. (\ref{defns}) and (\ref{defr}).
\end{itemize}
The same procedure can be carried out numerically if no analytical solution exists for $\theta_{\rm f}$ and $\theta_{\rm i}$. The details of the procedure are given in Ref. [\citen{Bilic:2018uqx}].

If $\varepsilon_{2}$ can be expressed as a function of $\varepsilon_{1}$ only, there exists a simplified way to find $n_{\rm s}$ and $r$ analytically based on solving the differential equation
\begin{equation}
\varepsilon_{2}=\frac{1}{\varepsilon_{1}}\frac{d\varepsilon_{1}}{dN},
\label{e2N}
\end{equation}
obtained from (\ref{slparametersdef}).
Integrating (\ref{e2N}) yields $\varepsilon_{1}$ as a function of $N$, with the integration constant fixed using a condition at the end of inflation. Then, we can plug $\varepsilon_{1}$ and $\varepsilon_{2}$ in Eqs. (\ref{defns}) and (\ref{defr}) to obtain $n_{\rm s}$ and $r$, respectively. We will use this approach in our further calculations whenever applicable, to obtain $n_{\rm s}$ and $r$ in terms of the  parameters that appear in generating test Hubble rate and the number of e-folds $N_{*}$.

\section{Various functional forms of the Hubble rate}
\label{various}

As already mentioned the model of inflation with the tachyon field minimally coupled to gravity has been investigated in Ref. [\citen{Aghamohammadi:2014aca}], using Hamilton-Jacobi formalism, assuming typical functions for the Hubble rate: the exponential function ($H(\theta)\propto e^{n\theta}$) and the power-law function ($H(\theta)\propto \theta^n$). While they concluded, for some value of the model parameters, the predictions of the model were compatible with the observational data available at that time (Planck 2015 [\citen{Planck:2015sxf}] and BICEP2 [\citen{BICEP2:2014owc}]), it is not confirmed by the latest Planck results from 2018 [\citen{Planck:2018jri}]. Employing this idea, we consider the tachyonic inflation model in several alternative background scenarios.

In the following subsections, we apply Hamilton-Jacobi formalism using specific functional forms of the Hubble function with the property $H_{,\theta}<0$. We reproduce some results from Ref. [\citen{Aghamohammadi:2014aca}] to compare them with our results obtained by making use of alternative functional forms of the Hubble rate, which have not been considered in the literature for standard tachyon inflation. We check the compatibility of the result for $n_{\rm s}$ and $r$ with the latest Planck data.

From here onwards it will be convenient to replace $\theta$ and $H(\theta)$ by dimensionless quantities
$\vartheta$ and $h(\vartheta)$ defined as
\begin{equation}
\vartheta\equiv \frac{\theta}{\ell}, \quad\quad h(\vartheta) \equiv \ell H(\ell\vartheta)
\label{eq0001}
\end{equation}
where $\ell$ is an unspecified length scale.

\subsection{Exponential function of the power function}

\begin{figure}[ht]
\centering
\begin{subfigure}{0.45\textwidth}
\centering
\includegraphics[width=\linewidth]{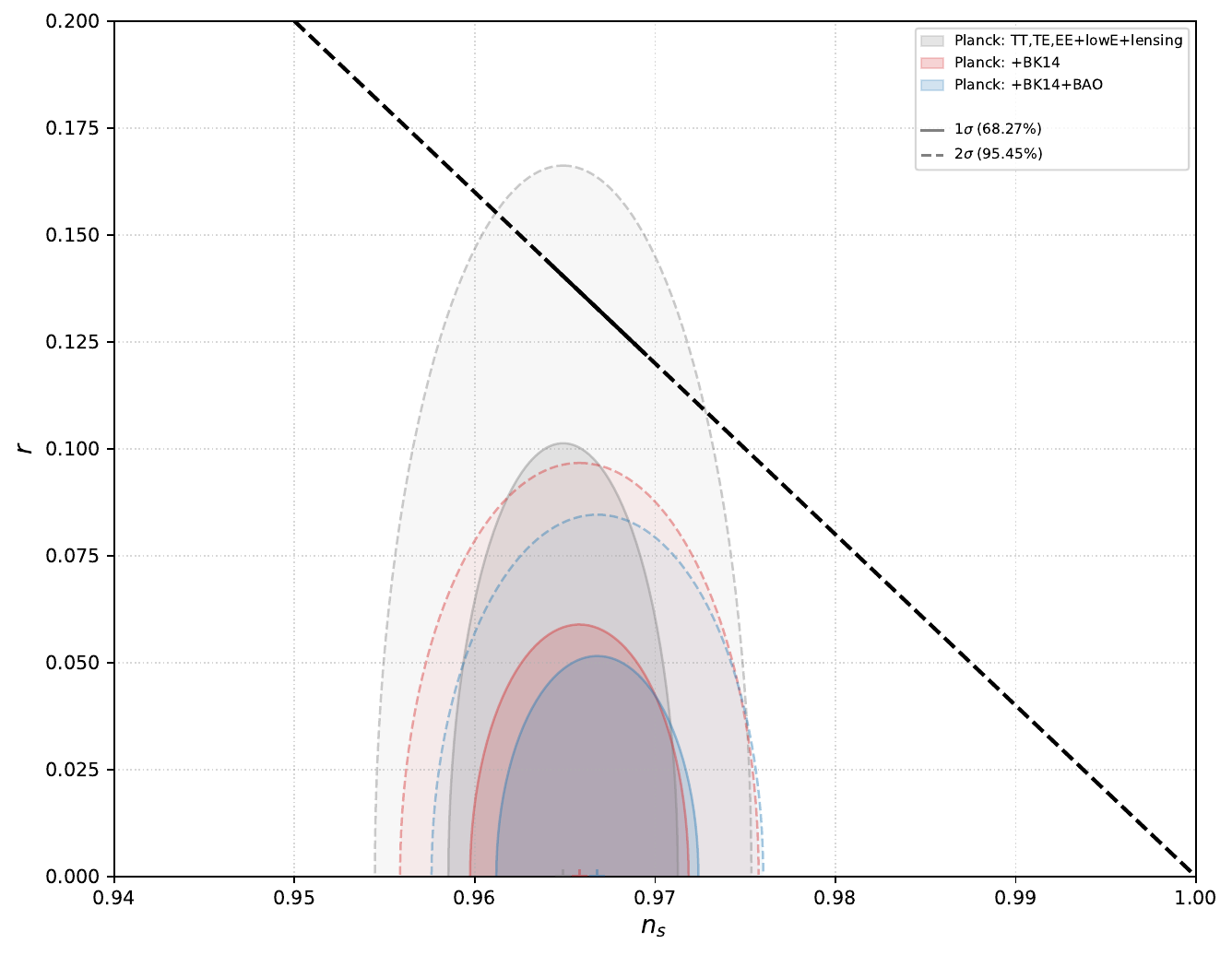}
\caption{Analytical results for the model $h(\vartheta)=\lambda e^{-n\vartheta}$ represented by the function $r=4-4n_{\rm s}$ (dashed line). The solid-line segment corresponds to the interval $55\leq N_{*}\leq 65$.\\\\\\}
\end{subfigure}
\hfill
\begin{subfigure}{0.45\textwidth}
\centering
\includegraphics[width=\linewidth]{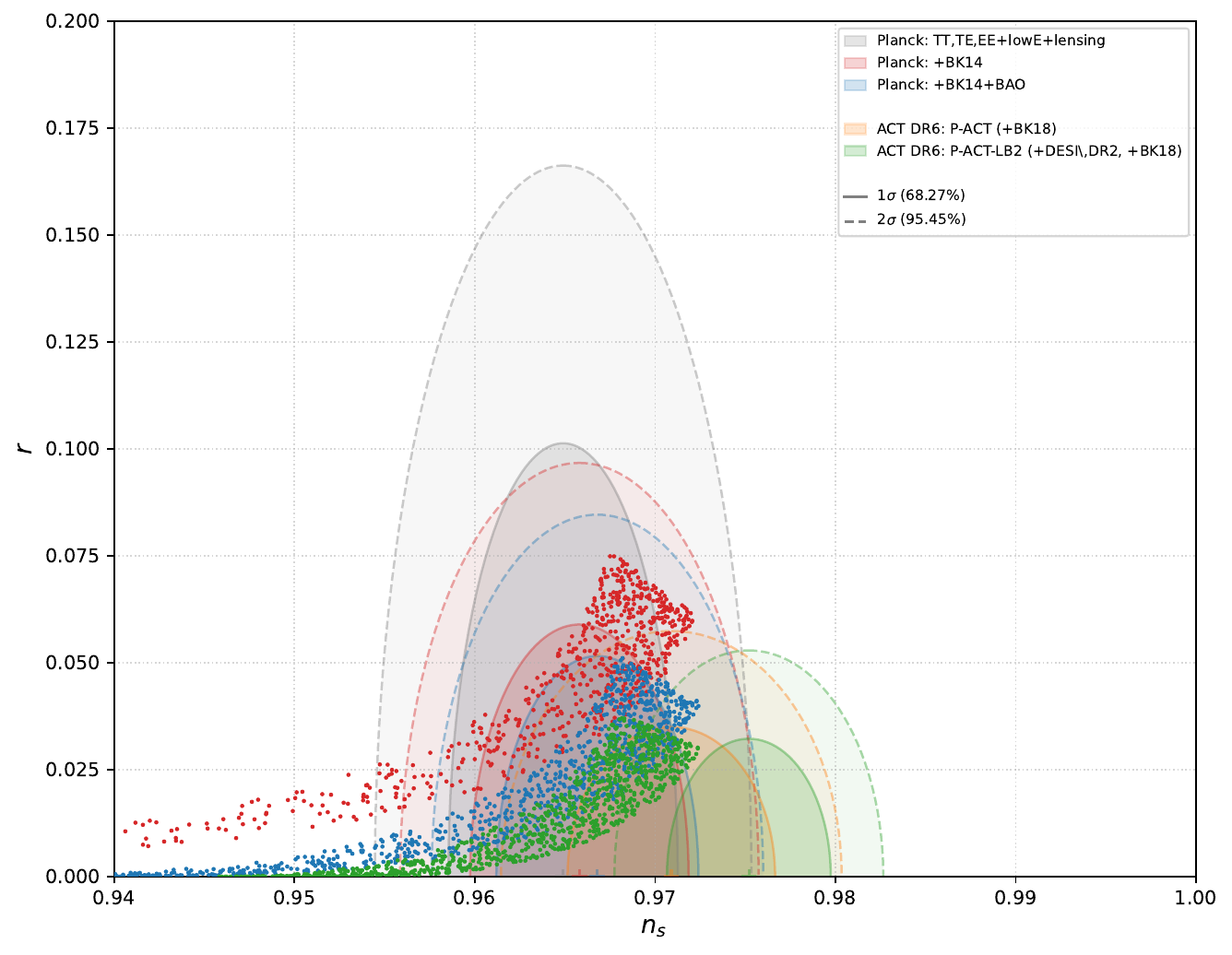}
\caption{Numerical results for the model $h(\vartheta)=\lambda e^{-\vartheta^n}$. The red, blue, and green dots correspond to $n=2$, $n=3$, and $n=4$, respectively. 
The parameter $\lambda$ varies in the interval $4\leq \lambda\leq 20$. The width of the stipes corresponds to the witdth of the  $N_*$ interval $55\leq N_{*}\leq 65$.\\}
\end{subfigure}
\caption{Theoretical predictions for the tensor to scalar ratio $r$ versus scalar spectral index $n_{\rm s}$ superimposed on observational constraints.}
\label{plotexp}
\end{figure}

We first reproduce the theoretical predictions of the Hubble rate as an exponential function
\begin{equation}
h(\vartheta)=\lambda e^{-n\vartheta},
\label{Hexpn1}
\end{equation}
initially studied in Ref. [\citen{Aghamohammadi:2014aca}],
where $\lambda $ and $n$ are free parameters.
We consider only $n>0$. In this case, the slow-roll parameters defined in (\ref{eqHJe1}) and (\ref{eqHJe2}), become
\begin{equation}
\varepsilon_{1}=\frac{2n^2e^{2n\vartheta}}{3\lambda ^2},
\end{equation}
\begin{equation}
\varepsilon_{2}=\frac{4n^2e^{2n\vartheta}}{3\lambda ^2}.
\end{equation}
According to (\ref{potential}), the expansion rate (\ref{Hexpn1}) yields the corresponding potential
\begin{equation}
V(\vartheta)= 3M_{\rm{Pl}}^2\ell^{-2}\lambda ^2e^{-2n\vartheta}\sqrt{1-\frac{4n^2e^{2n\vartheta}}{9\lambda ^2}}.
\end{equation}
Expressing $\varepsilon_{2}$ as a function of $\varepsilon_{1}$
\begin{equation}
\varepsilon_{2}=2 \varepsilon_{1},
\label{e1e2exp}
\end{equation}
allows one to integrate (\ref{e2N}) and find $\varepsilon_{1}$ as a function of the e-fold number $N$
\begin{equation}
\varepsilon_{1}=\frac{1}{1+2(N_{*}-N)}.
\label{e1exp}
\end{equation}
Here and in the following, we assume that $N=0$ at the beginning of inflation and inflation ends when $N = N_{*}$ where $\varepsilon_{1}(\theta_{\rm f})=1$. From (\ref{defns}) and (\ref{defr}), using (\ref{e1e2exp}) and (\ref{e1exp}), the observational parameters $n_{\rm s}$ and $r$ can be easily calculated for a given $N_{*}$. The results are depicted in the left panel in Fig. \ref{plotexp}. According to (\ref{e1exp}), $\varepsilon_{1}$ does not depend on the parameters $n$ and $\lambda$, yielding the plot $n_{\rm s}$ versus $r$ represented by the straight line defined by the function $r=4(1-n_{\rm s})$. It is evident that the number of e-folds in the interval $55\leq N_{*}\leq 65$, yields model predictions inconsistent with observational constraints. Therefore, one may conclude that the model with the Hubble (\ref{Hexpn1}), cannot successfully describe inflation.

Several attempts have been made to achieve a better agreement with observational constraints by considering a form slightly different from (\ref{Hexpn1}).
For example, in  Ref. [\citen{Videla:2016ypa}], the Hubble rate with a quasi-exponential dependence on the field has been proposed. Here, we propose  an exponential function of a power, i.e., 
\begin{equation}
h(\vartheta)=\lambda e^{-\vartheta^n},
\label{Hexpn2}
\end{equation}
where $n>1$, yielding
\begin{equation}
\varepsilon_{1}=\frac{2n^2e^{2\vartheta^n}\vartheta^{2(n-1)}}{3\lambda ^2},
\end{equation}
and
\begin{equation}
\varepsilon_{2}=\frac{4ne^{2\vartheta^n}\vartheta^{n-2}(n-1+n\vartheta^n)}{3\lambda ^2}.
\end{equation}
Combining (\ref{potential}) and (\ref{Hexpn2}), we find a potential 
\begin{equation}
V(\vartheta)= 3M_{\rm Pl}^2\ell^{-2}\lambda^2e^{-2\vartheta^n}\sqrt{1-\frac{4n^2e^{2\vartheta^n}}{9\lambda ^2}}.
\end{equation}
Now, the parameter $\varepsilon_{2}$ cannot be expressed as a function of  $\varepsilon_{1}$ to find find $\varepsilon_{1}$ as a function of $N$ using (\ref{e2N}). In this case, we calculate the parameters $n_{\rm s}$ and $r$ numerically, using the procedure described in section \ref{sec4}. The results are shown in the right panel in Fig. \ref{plotexp}. Clearly, for $55\leq N_{*}\leq 65$, the model $h(\vartheta) =\lambda e^{-\vartheta^n}$ is closer to a satisfactory fit of the observational data than the model with $h(\vartheta)=\lambda e^{-n\vartheta}$.  As shown in Fig.\ \ref{plotexp}, increasing the value of $n$ improves the agreement.

\subsection{Power-law  and modified power-law functions}

\begin{figure}[ht]
    \centering
    \begin{subfigure}{0.45\textwidth}
        \centering
        \includegraphics[width=\linewidth]{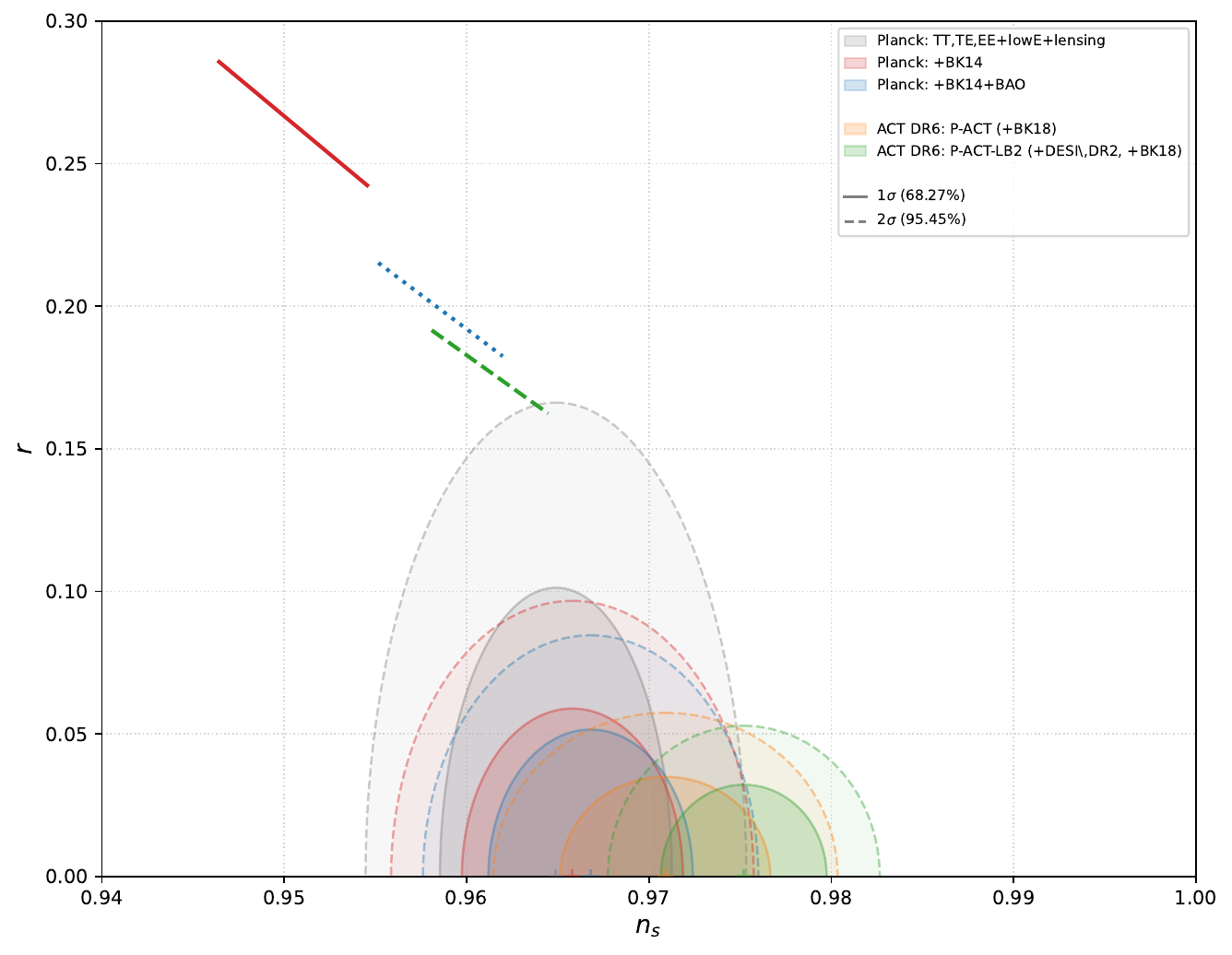}
\caption{Analytical results for the model  $h(\vartheta)=\lambda \vartheta^{-n}$ with $n=2$ (full red line), $n=3$ (dotted blue), and $n=4$ (dashed green).}  
    \end{subfigure}
    \hfill
    \begin{subfigure}{0.45\textwidth}
        \centering
        \includegraphics[width=\linewidth]{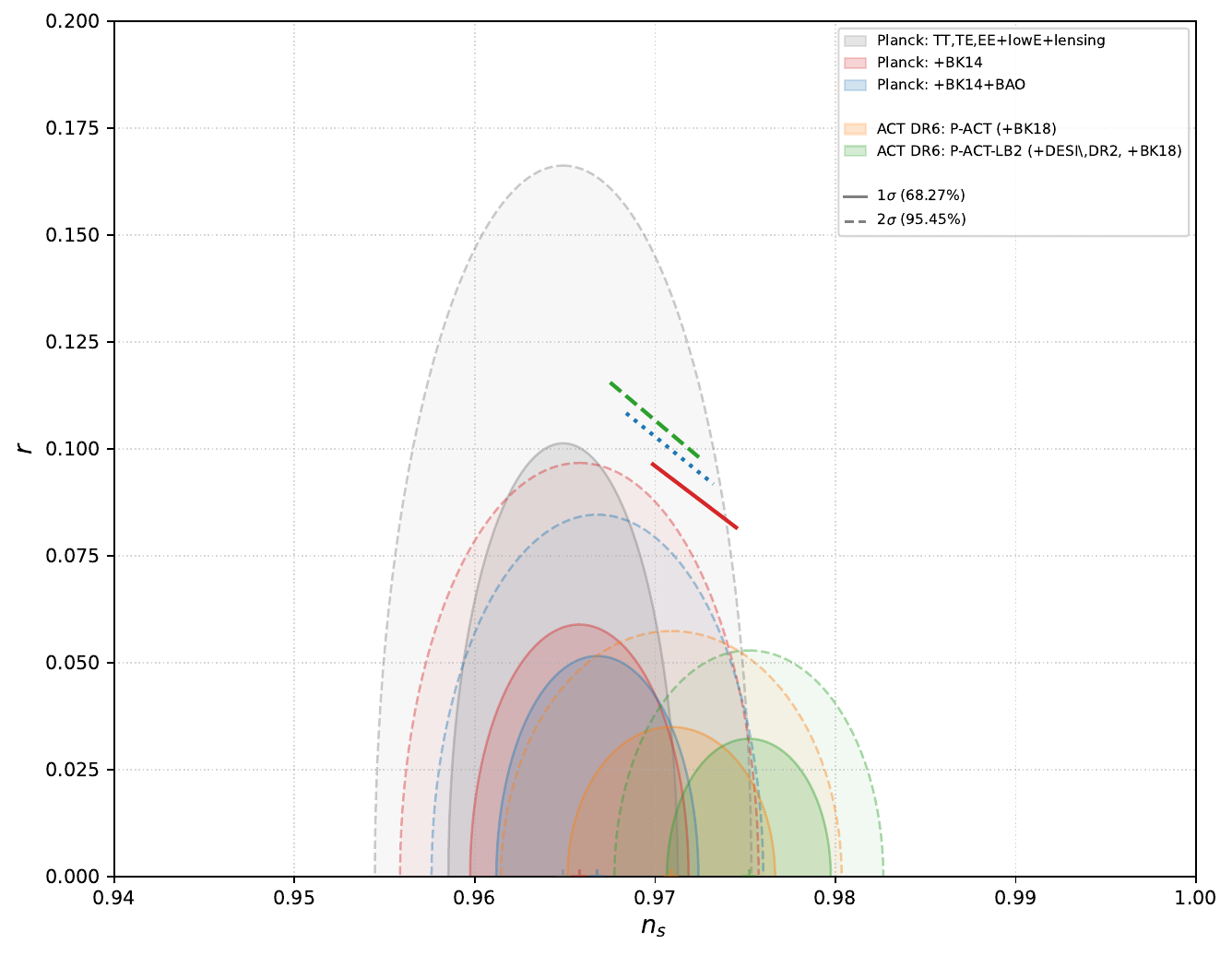}
        \caption{Analytical results for the model $h(\vartheta)=(\lambda-\vartheta)^n$  with $n=2$ (full red line), $n=3$(dotted blue), and $n=4$ (dashed green).} 
     \end{subfigure}
    \caption{Theoretical predictions for $r$ versus  $n_{\rm s}$ superimposed on observational constraints. The number of e-folds varies within the interval $55\leq N_{*}\leq 65$.}
    \label{figpowerlaw}
\end{figure}

The Hubble rate as a power-law function 
\begin{equation}
h(\vartheta)=\lambda \vartheta^{-n},
\label{Hpowertheta|}
\end{equation}
was also examined in Ref. [\citen{Aghamohammadi:2014aca}]. The first two slow-roll parameters 
\begin{equation}
\varepsilon_{1}= \frac{2n^2\vartheta^{2(n-1)}}{3\lambda^2},
\end{equation}
\begin{equation}
\varepsilon_{2}= \frac{4n(n-1)\vartheta^{2(n-1)}}{3\lambda^2},
\end{equation}
imply
\begin{equation}
\varepsilon_{2}=\frac{2(n-1)}{n} \varepsilon_{1}.
\end{equation}
Clearly, the choice $n=1$ is excluded since a constant $\varepsilon_{1}$ never reaches unity and the inflation stage never ends. For $n>1$, it is possible to find an analytical expression for  $\varepsilon _{1}$ as a function of the  e-fold number $N$
\begin{equation}
\varepsilon_{1}=\frac{1}{1+\frac{2(n-1)}{n}(N_{*}-N)}.
\label{e1Nexp}
\end{equation}
The potential in this case is
\begin{equation}
V(\vartheta)=3M_{\rm Pl}^2\ell^{-2}\lambda^2\vartheta^{-2n}\sqrt{1-\frac{4n^2\vartheta^{2(n-1)}}{9\lambda^2}}.
\end{equation}
The results for $n_{\rm s}$ and $r$ are shown in the left panel of Fig. \ref{figpowerlaw}.  Evidently, the Hubble rate $h(\vartheta)=\lambda \vartheta^{-n}$ cannot provide a reasonable agreement with observational data. 

Next,  assume the Hubble rate given by
\begin{equation}
h(\vartheta)=(\lambda -\vartheta)^n.
\label{powmod}
\end{equation}
A straightforward calculation gives 
\begin{equation}
\varepsilon_{1}=\frac{2}{3}n^2(\lambda -\vartheta)^{-2(1+n)},
\end{equation}
\begin{equation}
\varepsilon_{2}=\frac{4}{3}n(n+1)(\lambda -\vartheta)^{-2(1+n)},
\end{equation}
and 
\begin{equation}
\varepsilon_{2}=\frac{2(n+1)}{n} \varepsilon_{1}.
\label{eq0003}
\end{equation}
From (\ref{potential}), it follows
\begin{equation}
V= 3M_{\rm Pl}^2\ell^{-2}\sqrt{1-\frac{4n^2(\lambda -\vartheta)^{-2(n+1)}}{9}}(\lambda -\vartheta)^{2n}.
\label{Vrec1}
\end{equation}
In the slow-roll regime ($\varepsilon_{1}\ll 1$), the expression (\ref{Vrec1}) for $n=1$ reduces to 
\begin{equation}
V\simeq 3M_{\rm Pl}^2\ell^{-2}(\lambda -\vartheta)^{2}.
\end{equation}
This potential has already been studied in Refs. [\citen{Frolov:2002rr,Padmanabhan:2002cp}]. Using (\ref{e2N}) and (\ref{eq0003}) , we obtain  $\varepsilon _{1}$ as a function of the e-fold number $N$, similar to (\ref{e1Nexp})
\begin{equation}
\varepsilon_{1}=\frac{1}{1+\frac{2(n+1)}{n}(N_{*}-N)}.
\end{equation}
The right panel in Fig. \ref{figpowerlaw} shows that the obtained results for the model $h(\vartheta)=(\lambda -\vartheta)^n$ are slightly outside the range of the observational constraints, although closer than those of the model $h(\vartheta)=\lambda \vartheta^{-n}$.

\subsection{Powers of hyperbolic functions}
\label{hyper}

\begin{figure}[h!]
    \centering
    \begin{subfigure}{0.45\textwidth}
        \centering
        \includegraphics[width=\linewidth]{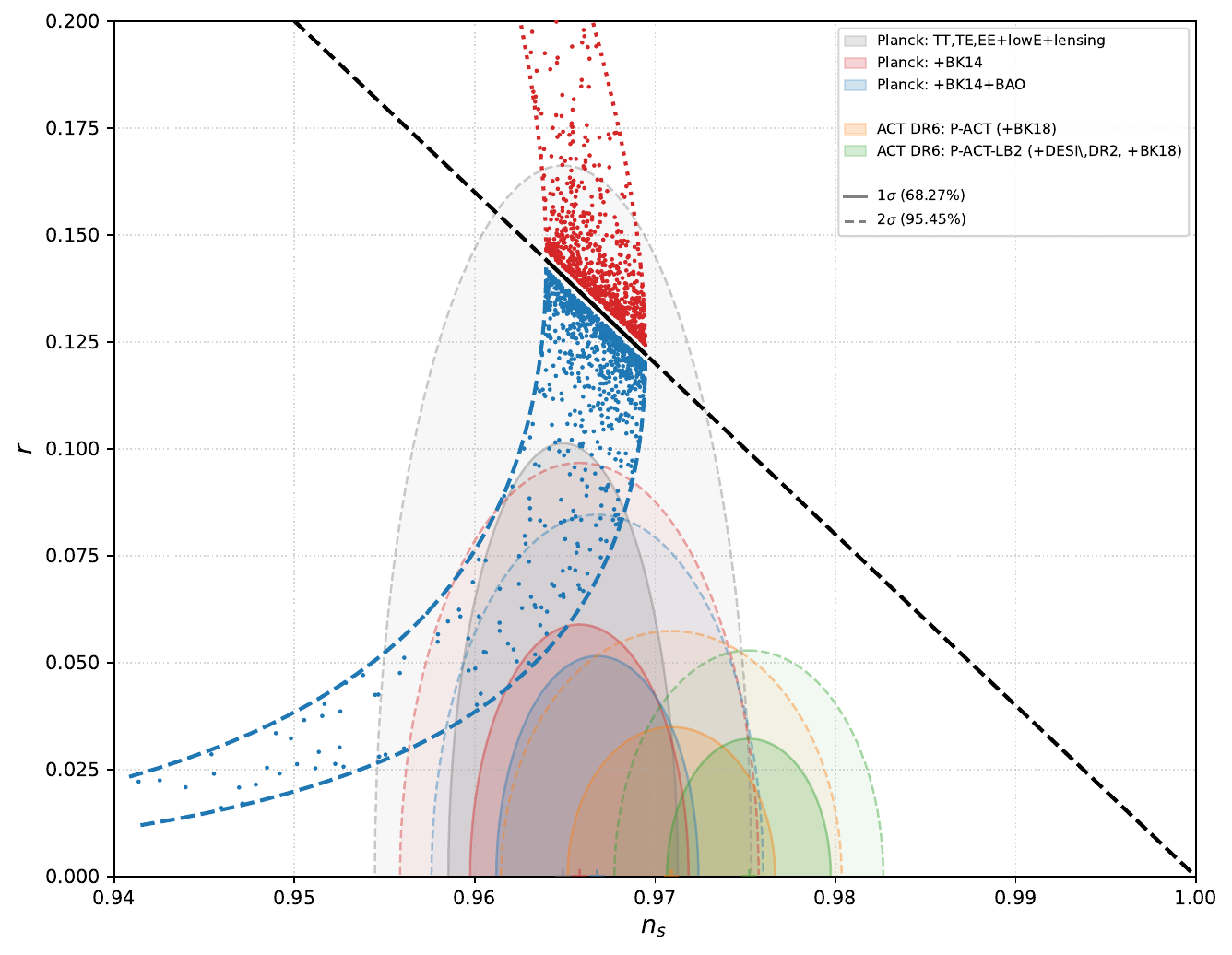}
        \caption{Numerical results for the models $h(\vartheta)=\lambda\sinh^{-1}\vartheta$ (blue dots) and $h(\vartheta)=\lambda\cosh^{-1}\vartheta$ (red  dots) for $\lambda$ variing in the interval $0.1\leq \lambda\leq 50$. The 
e-fold number varies between $N_{*}=55$ (dotted line) and $N_{*}=65$
(dashed line). The straight dahsed line represents the model $h(\vartheta)= \lambda e^{-n\vartheta}$  as in the left panel of Fig.~\ref{plotexp}.}
    \end{subfigure}
    \hfill
    \begin{subfigure}{0.45\textwidth}
        \centering
        \includegraphics[width=\linewidth]{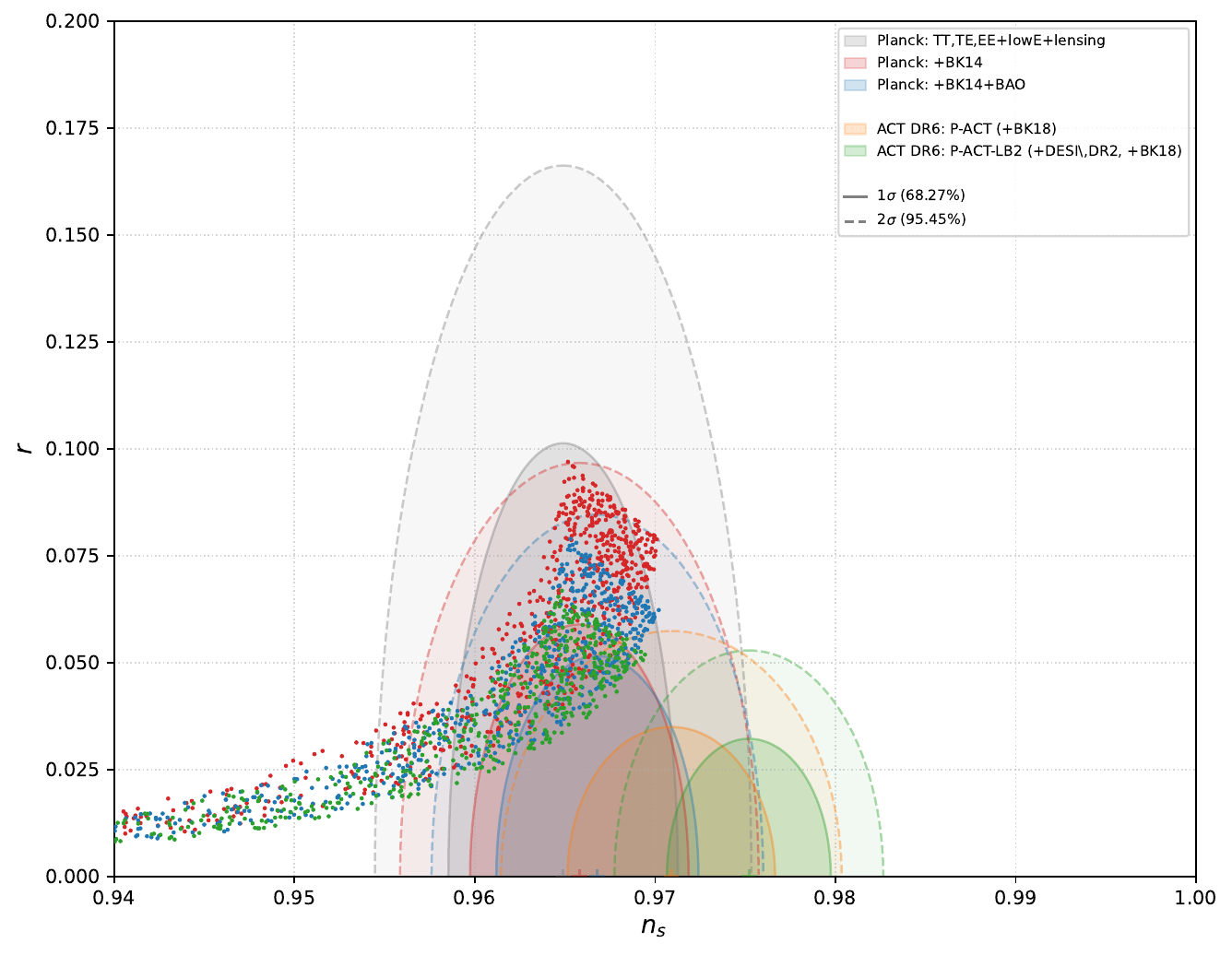}
        \caption{Numerical results for the model $h(\vartheta)=\lambda\cosh^{-n}\vartheta$  for $n=2$, 3, and 4, represented by red, blue, and green dots, respectively. 
The constant $\lambda$ varies  in the interval $0.1\leq \lambda\leq 20$.\\\\\\}
    \end{subfigure}
    \caption{Theoretical predictions superimposed on observational constraints for the models in which the Hubble rate is expressed in terms of the hyperbolic functions. 
The e-fold number varies in the interval $55\leq N_{*}\leq 65$.}
\label{fighyp}
\end{figure}

In this section, we investigate inflationary models in which the Hubble rate is expressed in terms of powers of the hyperbolic sine and cosine functions.
Consider first the Hubble rate as a fraction with  $\sinh\vartheta$ in the denominator, i.e.,
\begin{equation}
h(\vartheta)=\frac{\lambda }{\sinh\vartheta }.
\label{Hsinh}
\end{equation}
We find
\begin{equation}
\varepsilon_{1}=\frac{2\cosh^2\vartheta}{3\lambda^2},
\end{equation}
\begin{equation}
\varepsilon_{2}=\frac{4\sinh^2\vartheta}{3\lambda^2},
\end{equation}
and
\begin{equation}
\varepsilon_{2}=2\varepsilon_{1}-\frac{4}{3\lambda^2}.
\label{e1e2sinh}
\end{equation}
Plugging (\ref{Hsinh}) in (\ref{potential}) directly, we obtain the approximated potential in the slow-roll regime  
\begin{equation}
V\simeq 3M_{\rm Pl}^2\ell^{-2}\frac{\lambda^2 }{\sinh^2(\vartheta)}. 
\end{equation}
This model was considered in Ref.\ [\citen{G:2015vrt}]. Then, we find $\varepsilon _{1}$ as a function of $N$
\begin{equation}
\varepsilon_{1}=\frac{2}{(2-3\lambda^2)e^{-\frac{4}{3\lambda^2}(N_{*}-N)}+3\lambda^2}.
\end{equation}
Unlike previously, the parameter $\varepsilon_{1}$  is here affected by the parameter $\lambda$. The resalts presented in the left panel of  Fig. \ \ref{fighyp} show that the predictions of the model
 (\ref{Hsinh}) lie outside the region favored by the observations.

Next, we consider 
\begin{equation}
h(\vartheta)=\frac{\lambda}{\cosh\vartheta}.
\label{Hcoshn1}
\end{equation}
In this case, the expressions for $\varepsilon_{1}$ and $\varepsilon_{2}$ become
\begin{equation}
\varepsilon_{1}=\frac{2\sinh^2\vartheta}{3\lambda^2},
\end{equation}
\begin{equation}
\varepsilon_{2}=\frac{4\cosh^2\vartheta}{3\lambda^2},
\end{equation}
Then, we find
\begin{equation}
\varepsilon_{2}=2\varepsilon_{1}+\frac{4}{3\lambda^2},
\label{e1e2cosh}
\end{equation}
and
\begin{equation}
\varepsilon_{1}=\frac{2}{(2+3\lambda^2)e^{\frac{4}{3\lambda^2}(N_{*}-N)}-3\lambda^2}.
\end{equation}
The derived potential      
\begin{equation}
V\simeq 3M_{\rm Pl}^2\ell^{-2}\frac{\lambda^2 }{\cosh^2\vartheta}, 
\end{equation}
belongs to the class of potentials inspired by string-theory realizations [\citen{Barbosa-Cendejas:2017pbo}]. This potential has been studied in Ref.  [\citen{Mohammadi:2023sqy}].
Compared with the model (\ref{Hsinh}), the predictions of the model (\ref{Hcoshn1}) shown in Fig.\ \ref{fighyp}  are much better, though not quite satisfactory.

Next, consider  the Hubble rate as a fraction with the $n$-th power of  $\cosh^n\vartheta$ in the denominator, namely   
\begin{equation}
h(\vartheta)=\frac{\lambda}{\cosh^n\vartheta}.
\label{Hcoshn2}
\end{equation}
with  $n>1$.
Then, the Hubble flow parameters are given by  
\begin{equation}
\varepsilon_{1}=\frac{2n^2\cosh^{2(n-1)}\vartheta\sinh^2\vartheta}{3\lambda^2},
\label{e1coshn2}
\end{equation}
\begin{equation}
\varepsilon_{2}= \frac{2n\cosh^{2(n-1)}\vartheta(2-n+n\cosh(2\vartheta))}{3\lambda^2}.
\label{e2coshn2}
\end{equation}
In the slow-roll regime, the potential is approximated by
\begin{equation}
V\simeq 3M_{\rm Pl}^2\ell^{-2}\frac{\lambda^2}{\cosh^{2n}\vartheta}.
\end{equation}
In this case, since $\varepsilon_{2}$ cannot be expressed as an explicit function of  $\varepsilon_{1}$. Nevertheless, equations (\ref{e1coshn2}) and (\ref{e2coshn2}) 
represent  a function $\varepsilon_{2}=\varepsilon_{2}(\varepsilon_{1})$ in parametric form which we can treat numerically.
The predictions of the models (\ref{Hcoshn2})  with $n=1$, 2, 3, and 4,  are shown in the right panel of Fig. \ref{fighyp}.
The  $n_{\rm s}$ and $r$ predicted by the model are consistent with the bounds imposed by Planck 2018. 
Similar to the model (\ref{Hexpn2}), the agreement improves with increasing $n$.  
Note  that the difference between  expressions (\ref{e1e2exp}), (\ref{e1e2sinh}), and (\ref{e1e2cosh}) vanishes for large $\lambda$. 
This explains why the red and blue dots displayed in the left panel of Fig. \ref{fighyp} converge toward the straight line.

\begin{figure}[ht]
    \centering
    \begin{subfigure}{0.45\textwidth}
        \centering
        \includegraphics[width=\linewidth]{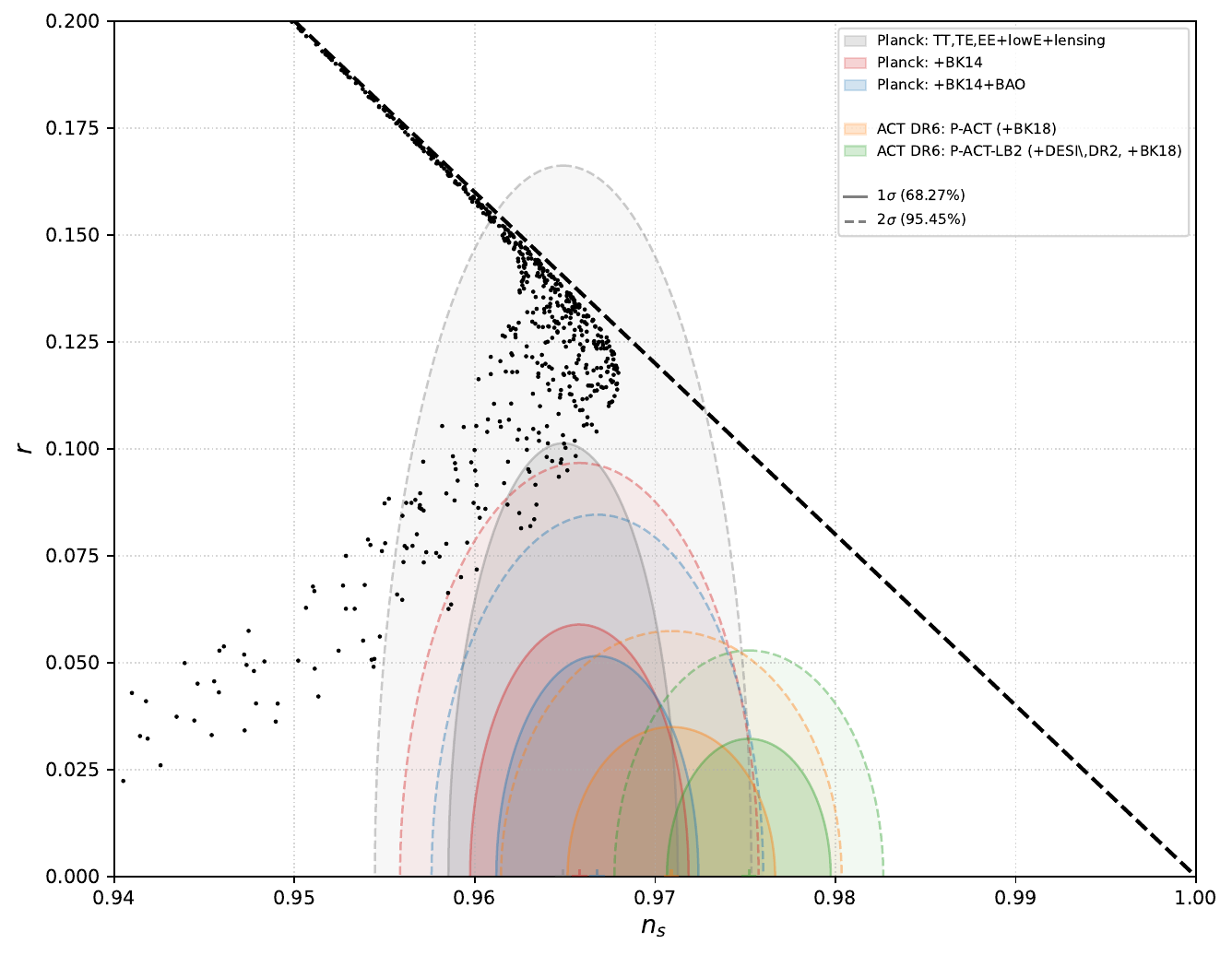}
        \caption{Numerical results obtained using equation (\ref{e1e2}).\\\\}
    \end{subfigure}
    \hfill
    \begin{subfigure}{0.45\textwidth}
        \centering
        \includegraphics[width=\linewidth]{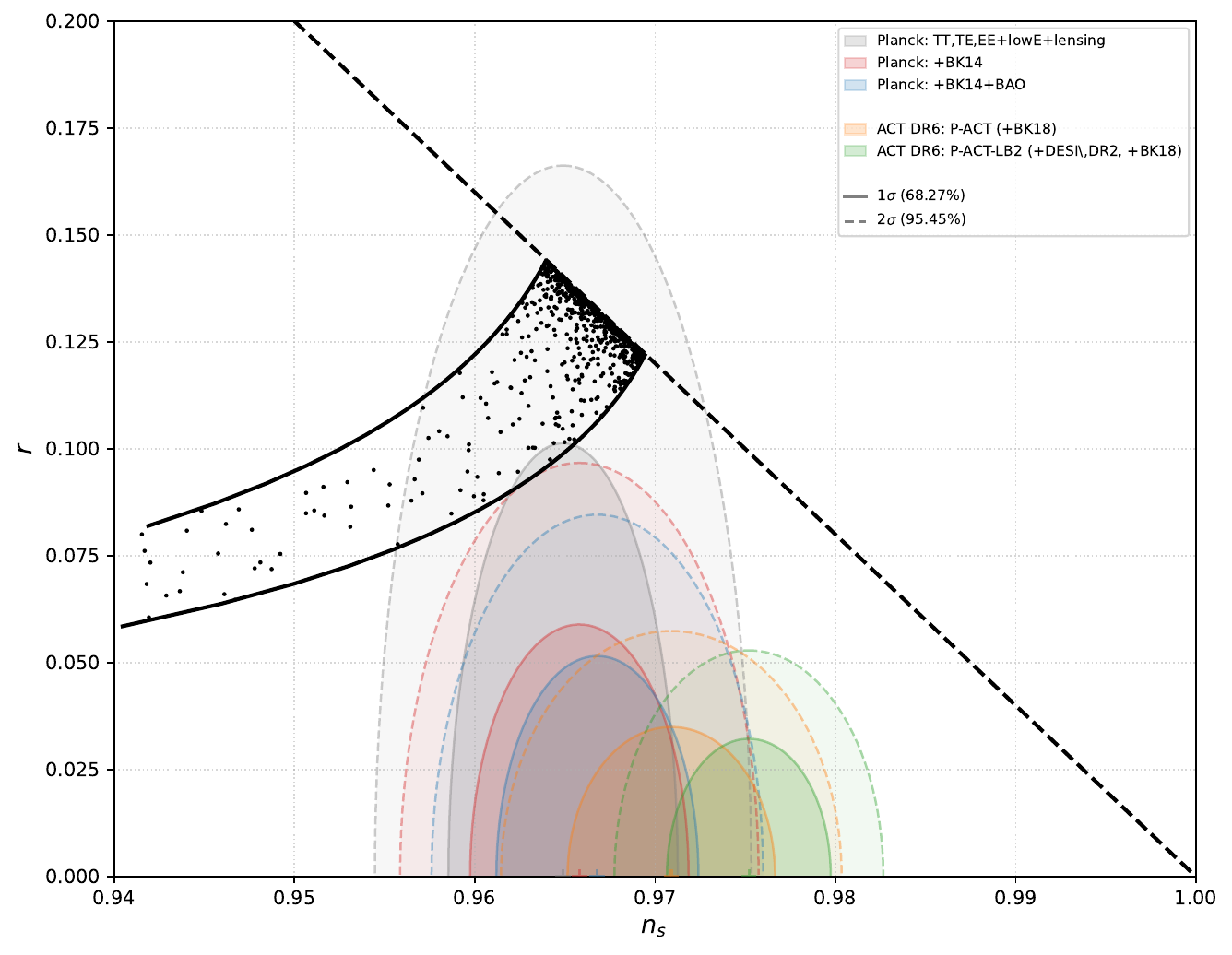}
        \caption{The results  obtained using equation (\ref{e1Ncos12app}). Dots represent numerical results, while solid lines represent analytical results.}
    \end{subfigure}
    \caption{Theoretical predictions superimposed on observational constraints for the model with $h(\vartheta)=\lambda\cosh^{-1/2}\vartheta$.
The number of e-folds varies in the interval $55\leq N_{*}\leq 65$. The straight dahsed line in both panels represents the model $h(\vartheta)= \lambda e^{-n\vartheta}$  as in the left panel of Fig.~\ref{plotexp}.}
\label{figcosh12}
\end{figure}

A particularly interesting model  arises for $n = 1/2$, i.e.,
\begin{equation}
h(\vartheta)=\frac{\lambda}{\cosh^{1/2}\vartheta},
\end{equation}
which, in the slow-roll regime, corresponds to the potential approximated by
\begin{equation}
V(\vartheta)=\frac{V_{0}}{\cosh\vartheta},
\end{equation}
where $V_{0}=M_{\rm Pl}^2\ell^{-2}\lambda^2$.
One obtains
\begin{equation}
\varepsilon_{1}=\frac{\sinh\vartheta\tanh\vartheta}{6\lambda^2},
\end{equation}
\begin{equation}
\varepsilon_{2}= \frac{(3+\cosh(2\vartheta))\sech\vartheta}{6\lambda^2},
\end{equation}
\begin{equation}
\varepsilon_{2}=\sqrt{4\varepsilon_{1}^{2}+\frac{4}{9 \lambda^4}}.
\label{e1e2}
\end{equation}
For a sufficiently large $\lambda$ , one finds   
\begin{equation}
\varepsilon_{2}\simeq 2\varepsilon_{1}+\frac{1}{9\lambda^4\varepsilon_{1}}.
\label{e1e2epprox}
\end{equation}
As shown in the left panel of Fig. \ref{figcosh12}, in the limit $\lambda \to \infty$ ( $\varepsilon_{2}\simeq 2 \varepsilon_{1}$), the results for $n_{\rm s}$ and $r$ tend to the line found in the model with an exponential function ($h(\vartheta)=\lambda e^{-n\vartheta}$). As expected, the results agree with those obtained in the tachyon model of Ref. [\citen{Steer:2003yu}].
The approximate expression (\ref{e1e2epprox}) yields  
 \begin{equation}
\varepsilon_{1}(N)=\frac{1}{3\sqrt{2}\lambda^{2}}\tan\left( \frac{\sqrt{2}}{3C_{1}^{2}}(N-N_{*})+\arctan(3\sqrt{2}\lambda^2) \right),
\label{e1Ncos12app}
\end{equation}
 which enables the analytical calculation of  $n_{\rm s}$ and $r$. 
The analytical results are depicted in the right panel of Fig.\ \ref{figcosh12}.

\section{Generating new functions for $H(\theta)$\label{sectionmethod}}
\label{method}
Up until now, we have analyzed different models, assuming specific forms of the function $H(\theta)$. Here, we introduce a method for generating new functional expressions for the Hubble rate, with the aim of analytical calculations of observational parameters. Starting from an assumed mathematical relationship between the slow-roll parameters, we construct the corresponding inflationary dynamics, thereby determining how inflation evolves and identifying the class of potentials that produce desired inflation.

Motivated by the results from the previous sections, such as the expressions (\ref{e1e2exp}), (\ref{e1e2sinh}), and (\ref{e1e2cosh}), we will assume that  the  parameter $\varepsilon_{2}$ can be expressed as a linear function of  $\varepsilon_{1}$, to wit,
\begin{equation}
\varepsilon_{2}=\alpha\varepsilon_{1}+\beta,
\label{e2ofe11}
\end{equation}
where $\alpha$ and $\beta$ are constant parameters. The parametrization (\ref{e2ofe11}), also proposed in Ref. [\citen{Yi:2018gse}], is supported by the  first-order  slow-roll  approximation [\citen{Liddle:1994dx}]. 

By setting $\alpha=0$ and $\beta=0$, one has  $\varepsilon_{2}=0$ corresponding to power-law inflation [\citen{Lucchin:1984yf}]. For certain values of  $\alpha$,
the parameters $\varepsilon_{2}$, $\eta\equiv -(\ddot{H})/(2H\dot{H})$, and $\eta_{H}\equiv (2H_{,\theta\theta})/(3H^3)$ take on constant values as required by 
the corresponding constant-roll scenarios.
Namely,  $\varepsilon_{2}=\beta$ for $\alpha = 0$, $\eta=\varepsilon_{1}-\varepsilon_{2}/2=-\beta/2$ for $\alpha = 2$,  and $\eta_{H}=2\varepsilon_{1}-\varepsilon_{2}/2=-\beta/2$ for  $\alpha = 4$. 

In section \ref{hyper}, linear relationships (\ref{e1e2sinh}) and (\ref{e1e2cosh}) have been found for the models $h(\vartheta)=\lambda\sinh^{-1}\vartheta$ and $h(\vartheta)=\lambda\cosh^{-1}\vartheta$, respectively.
As we can see from Fig.\ \ref{fighyp}, the latter model yields a better agreement with observations than the former, which suggests $\alpha>0$ and $\beta\geq 0$. Integrating (\ref{e2N}) together with (\ref{e2ofe11}) we obtain
\begin{equation}
\varepsilon_{1}=\frac{\beta}{(\alpha+\beta)e^{\beta(N_{*}-N)}-\alpha}.
\label{e1AB1} 
\end{equation}
The constant of integration is chosen such that $\varepsilon_{1}(N_{*})=1$. 
From (\ref{e1AB1}), for $\alpha=4$  and $\beta=-2\eta_{H}=\rm{const}$, we recover the result of Ref. [\citen{Gao:2018tdb}].
 
Applying equations (\ref{eqHJe1}) and (\ref{eqHJe2}) to  (\ref{e2ofe11}) we find a differential equation for $H$
\begin{equation}
H_{,\theta\theta}H-(2-\frac{\alpha}{2})H_{,\theta}^2+\frac{3}{4}\beta H^4=0.
\label{difeq1H}
\end{equation}
To test the method described above, we solve equation (\ref{difeq1H}) for fixed values of  $\alpha$ and $\beta$. The type of solution to equation (\ref{difeq1H}) is the most affected by the value of $\alpha$.  For $\alpha=1$, we find an analytical solution
\begin{equation}
h(\vartheta)=\frac{4 C_{1}}{C_{1}^2(\vartheta+C_{2})^2+6\beta},
\label{solA1}
\end{equation}
where $C_{1}$ and $C_{2}$ are an arbitrary dimensionless constants. Given $h(\theta)$, integrating Eq. (\ref{HJ2}) yields $\vartheta$ as a function of $t$. Then,
 according to Eq. (\ref{sf}), the expansion scale is given by
\begin{equation}
a(t)=a_{0}\exp\left(\frac{1}{3\beta}(2C_{1}\tau-3\log(C_{1}^2 e^{C_{1}(\frac{2\tau}{3}+C_{3})})+6\beta)\right).
\end{equation}
where we have introduced a dimensionless time variable $\tau=t/\ell$.

For $\alpha=2$, Eq. (\ref{difeq1H}) admits two solutions
\begin{equation}
h(\vartheta)=\frac{4C_{1}e^{\sqrt{C_{1}}(C_{2}+\vartheta)}}{3\beta C_{1}+e^{2\sqrt{C_{1}}(C_{2}+\vartheta)}},
\label{solA2}
\end{equation}
and 
\begin{equation}
h(\vartheta)=\frac{4C_{1}e^{\sqrt{C_{1}}(C_{2}+\vartheta)}}{1+3\beta C_{1}e^{2\sqrt{C_{1}}(C_{2}+\vartheta)}}.
\end{equation}
Both solutions yield the same expansion scale 
\begin{equation}
a(t)=a_{0}\cosh^{-2/\beta}\left(\frac{\sqrt{\beta C_{1}}(3C_{3}+2\tau)}{2\sqrt{3}}\right).
\end{equation}
Note that the solution (\ref{solA2}) coincides with the expression for the Hubble rate in a tachyon-driven inflation model with a constant negative slow-roll parameter $\eta$ [\citen{Stojanovic:2023qgm}]. 

For $\alpha>2$, the solutions to equation (\ref{difeq1H}) are inverse functions of the hypergeometric functions, and one  cannot go further analytically, except for $\alpha=4$, in which case
\begin{equation}
h(\vartheta)=\frac{2^{3/4}}{3^{1/4}}\left(\frac{B}{C_{1}}\right)^{1/4} \operatorname{sn}\left(\frac{3^{1/4}}{2^{3/4}}\left(\frac{B}{C_{1}}\right)^{1/4}(\vartheta+C_{2}) , -1\right),
\label{soljacobi}
\end{equation}
where the Jacobi elliptic function $\operatorname{sn}(u,-1)$ is the inverse of the function
\begin{equation}
 u(y)=\int_{0}^{y}\frac{dx}{\sqrt{1-x^4}}.
\end{equation}
In this case, one cannot analytically derive the function describing the evolution of the expansion scale.

The case $\beta=0$ must be treated separately. In this case, from (\ref{e2N}) and (\ref{e2ofe11})  we have 
\begin{equation}
\varepsilon_{1}=\frac{1}{1+\alpha(N_{*}-N)},
\label{e1AB2} 
\end{equation}
and  equation (\ref{difeq1H}) simplifies to
\begin{equation}
H_{,\theta\theta}H-(2-\frac{\alpha}{2})H_{,\theta}^2=0.
\label{difeq2H}
\end{equation}
For $\alpha\geq 1$ and $\alpha\neq 2$, its solution is of the form 
\begin{equation}
h(\vartheta)=C_{1}\left((\alpha-2)\vartheta-2C_{2}\right)^{\frac{2}{\alpha-2}},
\label{solB1}
\end{equation}
and yields the expansion scale 
\begin{equation}
a(t)=a_{0}\exp\left(
\frac{C_{1}(3C_{3}+2\tau)(\alpha-1)}{2\alpha}\left(2^{\frac{\alpha-2}{\alpha-1}}3^{\frac{\alpha-2}{2(\alpha-1)}}\left(-\frac{(3C_{3}+2\tau)(\alpha-1)}{C_{1}}\right)^{\frac{\alpha-2}{2(\alpha-1)}}
\right)^{\frac{2}{\alpha-2}}\right) .
\end{equation}
For $\alpha=2$
\begin{equation}
h(\vartheta)=C_{1}e^{C_{2}\vartheta },
\label{solB2}
\end{equation}
with the expansion scale  
\begin{equation}
a(t)=\exp\left(-\frac{1}{3}C_{2}^2(3C_{3}+\tau)\tau\right).
\end{equation}

It is worth checking that some of the assumed  Hubble rate functions in section \ref{various} are special cases of certain values of the coefficients in (\ref{solA2}), (\ref{solB1}), and (\ref{solB2}). 
The function $h(\vartheta)=\lambda e^{-n\vartheta}$ is obtained by substituting $C_{1}=\lambda$, $C_{2}=-n$ into (\ref{solB2}).
 
The functions $h(\vartheta)=\lambda \vartheta^{-n}$ and $h(\vartheta)=(\lambda -\vartheta)^n$ are obtained from (\ref{solB1}), with $C_{1}=\lambda (\lambda-2)^{-2/(\alpha-2)}$, $C_{2}=0$ and $\alpha=2(n-1)/n$, or with  $C_{1}=(-n/2)^{n}$, $C_{2}=\alpha/n$ and $\alpha=2(n+1)/n$.  The function $h(\vartheta)=\lambda\sinh^{-1}\vartheta$ is obtained by substituting $C_{1}=1$, $C_{2}=\ln (2/\lambda)$ and $\beta=-4/(3\lambda^2)$ into (\ref{solA2}). The same values $C_{1}$ and $C_{2}$ together with $\beta=4/(3\lambda^2)$ lead to $h(\vartheta)=\lambda\cosh^{-1}\vartheta$. It is important to stress that the functions  (\ref{solA1}) and (\ref{soljacobi}) have been derived from the assumption of a linear relationship between sow-roll parameters, rather than being arbitrarily assumed.

\begin{figure}[h!]
\begin{center}
\includegraphics[scale=0.5]{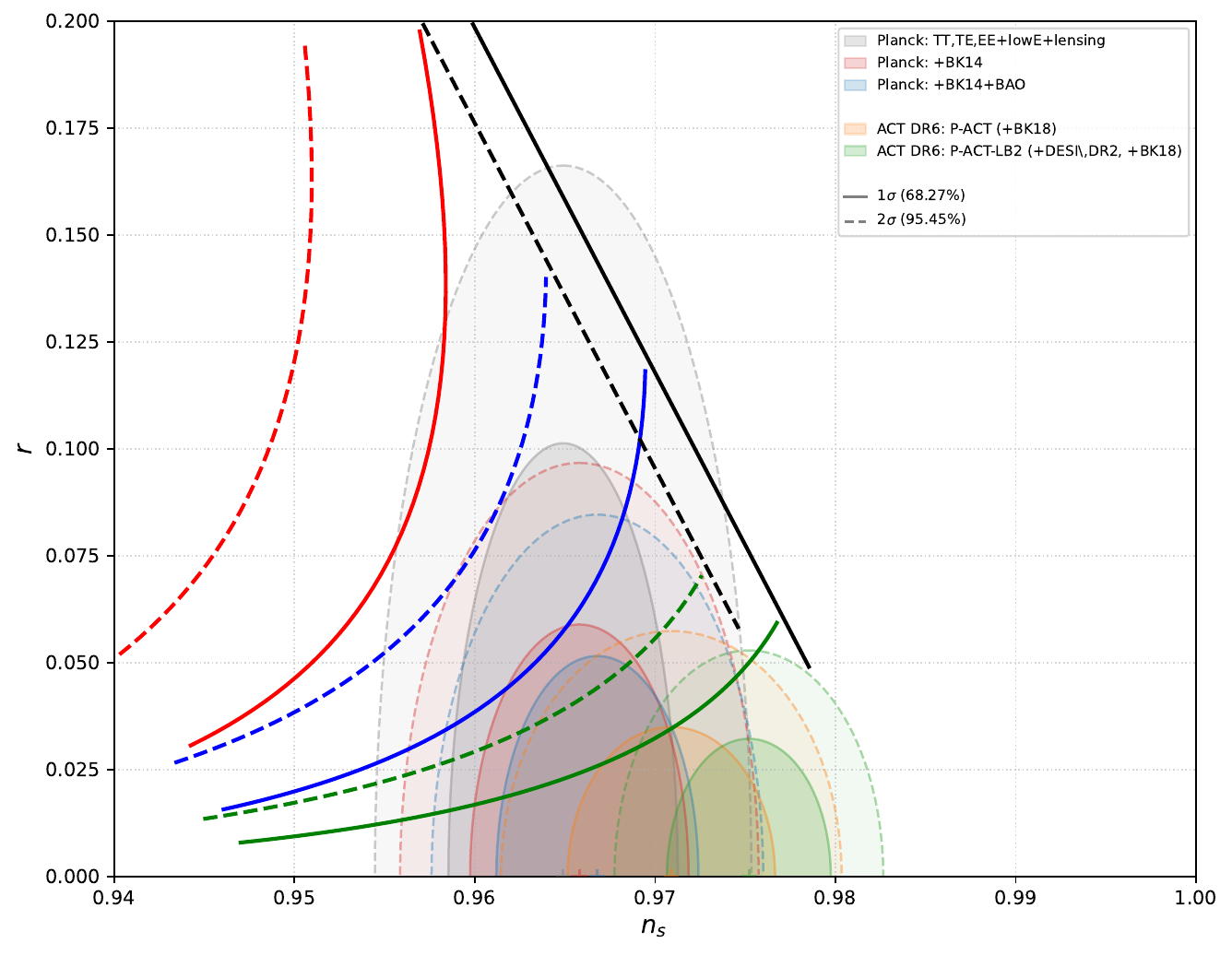}
\caption{Theoretical predictions superimposed on observational constraints for the model  (\ref{e1AB1}) with $\alpha=1$ (red  lines), $\alpha=2$ (blue lines), and $\alpha=4$ (green lines)  for $\beta$ ranging in the intertal $0.001\leq \beta\leq 0.5$, and for the model (\ref{e1AB2}) with  $\beta=0$ and $\alpha$ ranging between $1$ and $5$ (straight black lines). The number of e-folds is $N_{*}=55$ and $N_{*}=65$ for solid and dashed lines, respectively.}
\label{figcoupling}
\end{center}
\end{figure}

In Fig. \ref{figcoupling}, we depict theoretical predictions for the observational parameters for different values of the parameters $\alpha$ and $\beta$ that appear in the expressions  (\ref{e1AB1}) and (\ref{e1AB2}) derived analytically.
As shown in Fig.\ \ref{figcoupling}, an increase in the parameter $\alpha$,
systematically shifts the observational predictions toward higher values of $n_s$ and smaller values of $r$. Should upcoming combined data from Planck 2018, DESI, ACT, or the forthcoming Simons Observatory firmly establish a shift toward the lower-right region of the parameter space, this parametric behavior indicates that our framework could naturally accommodate such observations. Ultimately, this highlights that the proposed ansatz (\ref{e2ofe11}) and its modifications provide a flexible and promising foundation for future analytical and numerical considerations.

\section{Conclusion}
\label{conclude}

This paper has reconsidered the standard tachyon inflation using an analog of the Hamilton-Jacobi formalism. We stress that our work differs from [\citen{Aghamohammadi:2014aca}] in that it assumes a new form of the Hubble rate and the latest observational constraints. As a novelty, we have proposed the Hubble rate function as a modified power-law, an exponential of the power, the hyperbolic sine, the hyperbolic cosine, and their powers. For each function of the Hubble parameter, we have calculated the observational parameters $n_{\rm s} $ and $r$, and compared the model predictions with observational data for the number of e-fold in the range $55\leq N_{*}\leq 65$.

A minimally coupled tachyon field inflation with $h(\vartheta)=\lambda e^{-n\vartheta}$ and $h(\vartheta)=\lambda \vartheta^{-n}$ was supported by the Planck 2013 data but is ruled out by the latest Planck 2018 results. Assuming $h(\vartheta)=\lambda\cosh^{-1}\vartheta$, we have obtained values of $n_{\rm s}$ and $r$ that are in good agreement with observational data for the appropriate choice of the free parameter. We conclude that the obtained observational parameters from the models with $h(\vartheta)=\lambda\sinh^{-1}\vartheta$ and $h(\vartheta)=(\lambda-\vartheta)^n$ are ruled out by the observational data. We have improved the results for the observational parameters obtained for the Hubble rate as the exponential function and the hyperbolic cosine function, by taking their powers, i.e., $h(\vartheta)=\lambda e^{-\vartheta^n}$ and $h(\vartheta)=\lambda\cosh^{-n}\vartheta$. This has necessitated numerical calculations. It is worth noting that the reconstructed type of the tachyonic potential for the models with $h(\vartheta)=(\lambda-\vartheta)^n$, $h(\vartheta)=\lambda\sinh^{-1}\vartheta$ and $h(\vartheta)=\lambda\cosh^{-1}\vartheta$ has already appeared in string theory and were used previously in the literature for tachyon inflation.

In addition, we have proposed a general method to generate functions for $H(\theta)$ that enable analytical computations of $n_{\rm s}$ and $r$. Assuming a linear relationship between $\varepsilon_{2}$ and $\varepsilon_{1}$, we have found analytic expressions for the Hubble expansion rate for specific values of the coefficients. This procedure leads to a family of solutions to which all functions for the Hubble rate discussed in this paper belong. To the best of our knowledge, this approach has not been introduced previously. One significant advantage of this approach is that it avoids the need to introduce ad hoc functions for the Hubble rate. It would be interesting to extend the present analysis to alternative inflationary scenarios beyond the standard tachyon inflation, e.g., inflation in the braneworld scenarios [\citen{Stojanovic:2023dni,Bilic:2018uqx,Bilic:2016fgp}].

Notably, while our primary focus remains anchored on the Planck 2018 baseline, our framework inherently possesses the parametric flexibility to accommodate recent observational hints. In particular, the potential upward shift in the scalar spectral index $n_{\rm{s}}$ suggested by joint multi-probe analyses including DESI and ACT positions this approach as a viable candidate for upcoming cosmological datasets.
Furthermore, exploring a more generalized functional relation between hierarchical slow-roll parameters offers a promising foundation for future analytical and numerical investigations of observational parameters, including, potentially, the running of the spectral index within the framework of standard tachyon inflation, as well as related models previously considered and cited in this paper.

\section*{Acknowledgments}

This work has been supported by the ICTP--SEENET-MTP project NT-03 in 2025 and the COST Action CA23130 ``Bridging high and low energies in search of quantum gravity (BridgeQG)''. The work of G.S.Dj., D.D.D., and M.M. was supported by the Ministry of Science, Technological Development and Innovation of the Republic of Serbia under contract no.\ 451-03-34/2026-03/200124, while the work of M.S.\ was supported by contract no.\ 451-03-34/2026-03/200113. G.S.Dj. acknowledges the support of the CEEPUS projects RS-1514-05-2425 and M-RS-1514-2526, and the CERN-TH department, thanking the latter also for its kind hospitality and numerous useful consultations on the subject.


\end{document}